\documentclass[prl,twocolumn]{revtex4}

\usepackage{epsfig}
\usepackage{amssymb,amsmath,stmaryrd,tabularx}
\usepackage{bm}
\usepackage{graphicx}
\begin{document}
\title{Drying Patterns: Sensitivity to Residual Stresses}
\author{Yossi Cohen$^1$, Joachim Mathiesen$^2$, and Itamar Procaccia$^1$}
\begin{abstract}
Volume alteration in solid materials is a common cause of material failure. Here we investigate the crack formation in thin elastic layers attached to a substrate. We show that small variations in the volume contraction and substrate restraint can produce widely different crack patterns ranging from spirals to complex hierarchical networks. The networks are formed when there is no prevailing gradient in material contraction whereas spirals are formed in the presence of a radial gradient in the contraction of a thin elastic layer.
\end{abstract}
\affiliation{$^1$Department of Chemical Physics, The Weizmann Institute of Science, Rehovot 76100, Israel\\
$^2$Physics of Geological Processes, University of Oslo, Oslo, Norway }
\date{\today}
\maketitle

\begin{figure}
\epsfig{width=.42\textwidth,file=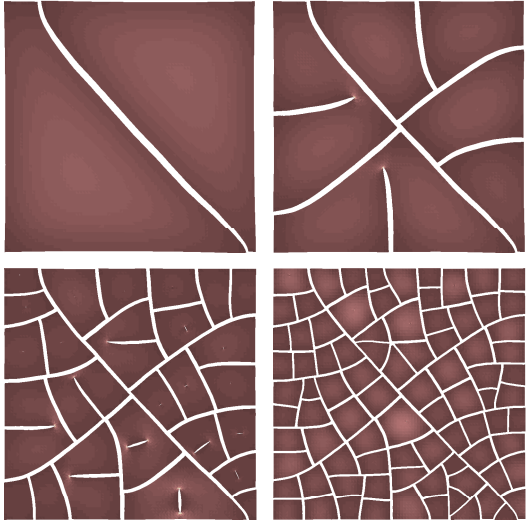}
\caption{Color online. Four different stages in the evolution of a hierarchical crack network. The total contraction was $9\%$ and cracks were nucleated inside domains whenever the maximum principal stress exceeded $\sigma_c=0.85$ in arbitrary units.  The substrate restraining force was drawn from a uniform distribution with $10\%$ disorder and a mean of unity. }
\label{distarea}
\end{figure}

{\bf Introduction}. Desiccation is known to produce complex networks of shrinkage-cracks in starch-water mixtures or clays\cite{88SM,94G,05B2,05BPAAC}. In concrete small cracks are often formed by the preparatory drying process and by the later ingress of reactive reagents. Similarly in nature, the infiltration of fluids and chemical reagents into rocks generate internal stresses that form intricate patterns of pervasive cracks\cite{08IJMMF}. Typically the stress is generated from local volume changes. Fractures are also observed in thin films attached to a substrate. Experiment on films have revealed intricate patterns ranging from the hierarchical structure typically observed in mud and concrete to spiral shaped cracks \cite{88SM, 03S}.  In spin-coating a fluid droplet is added at the center of a rotating substrate and is spread by centrifugal forces to cover the full substrate. During the drying and curing of the system, chemical bonds are formed between the coating and the substrate. In this process the coating often shrinks and tensile stresses are produced \cite{00M} and if one is less careful, the stress may result in an unwanted cracking of the coating. Due to the spinning of the system, the residual stress of the coating may also contain inherent shear components. Other mechanisms like anisotropic drying rates can also leave behind remnant shear. In cases where the thickness of the drying specimen is small and the contraction fairly uniform, i.e. no residual shear stresses, the growing cracks typically form an intricate hierarchical pattern. The pattern is the result of a cascade of successive cracks (which is supported by experimental evidences \cite{05BPAAC}); at each fragmentation stage, a crack is forming which divides a mother fragment of area A into two daughter fragments of areas $A_1$ and $A_2$ respectively, with area conservation ($A = A_1 + A_2$). It is worth noticing that in principle the trajectory of the crack that is dividing the mother fragment can be anything, and is determined from the shape and size of the mother domain and the inherent material disorder.

In order to analyze the spiral and hierarchical cracks, we consider a system consisting of a thin elastic layer attached to an elastic substrate. Under plane stress conditions we have that the in-plane strain tensor $\epsilon_{ij}$ in the layer is related to the stress tensor $\sigma_{ij}$ by
\begin{eqnarray}
\epsilon_{xx}&=&(\sigma_{xx}-\nu \sigma_{yy})/E {}+\beta\ , \nonumber \\
 \epsilon_{yy}&=&(\sigma_{yy}-\nu \sigma_{xx})/E {}+\beta\ , \nonumber\nonumber\\
\epsilon_{xy}&=&(1+\nu)\sigma_{xy}/E,\label{stressstrain}
\end{eqnarray}
where $E$ is Young's modulus and $\beta$ (in the absence of external stress) is a measure of the free volume change caused by e.g. drying or thermal expansion of the thin elastic layer. Whenever the film is displaced from its equilibrium position by a local displacement $u$ the elastic substrate tries to restore the film by a force $\mathbf f(\mathbf u)$. For small displacements we assume that this force is linearly proportional to $-\mathbf u$. In general it is assumed that volume alteration in the film happens on a time scale much larger than the time required for elastic waves to propagate across the system and the system is therefore assumed to always be in elastostatic equilibrium. The force balance therefore assumes the form
\begin{equation}
  \label{forcebalance}
  \partial_j  \sigma_{ij}-\mu u_i=0 \ ,
\end{equation}
where $\mu$ is the constant of proportionality of the substrate restoring force. For small deformations the strain follows from the displacement via the relations $\epsilon_{ij}=(\partial_j u_i +\partial_i u_j)/2$. Combining this relation with the force balance Eq. (\ref{forcebalance}) and stress-strain relations Eq. (\ref{stressstrain}) we achieve the following equation for the displacement
\begin{equation}
  \label{dis}
\triangle\mathbf u+\frac {1+\nu} {1-\nu}\mathbf \nabla (\nabla\cdot \mathbf u)=\frac {2(1+\nu)\mu}{E} \mathbf u \ .
\end{equation}

We now provide an estimate of the typical stress encountered during volume alteration of thin films with a linear spatial extend of size $R$. To that end, we shall consider the maximum stress for a circular domain of radius $R$ located at the center of coordinates and with vanishing stress at the boundaries. The displacement field is for a uniform material contraction $\beta$ found as a solution to the radial symmetric version of Eq. (\ref{dis})
\begin{equation}
\frac{\partial^2u_r}{\partial r^2} +\frac 1 r \frac{\partial u_r}{\partial r} -\left (a+\frac 1 {r^2}\right) u_r =0 \ ,
\label{pde}
\end{equation}
where the material specific constant $a$ is given by $a=(1-\nu^2){\mu}/{E}$.
Multiplying both sides of Eq. (\ref{pde}) by $r^2$ and rescaling $r$ with $\sqrt{a}$ yields the modified bessel differential equation. The solution for $\sigma_{rr}(R)=0$ and $u_r(0)=0$ is given by
\begin{equation}
  \label{circular}
  u_r(r)=\frac{\beta R I_1(\sqrt a r)}{\sqrt a R I_0(\sqrt a R)-(1-\nu)I_1(\sqrt a R)}, \quad u_{\theta}=0.
\end{equation}
Here $I_n$, $n=0,1$ is the modified bessel function of the first kind. From the displacement field, the stress in cylindrical coordinates follows from the expressions
\begin{eqnarray}
  \label{srr}
  \sigma_{rr}(r)&=&\frac E {1-\nu^2}\left(\frac{\partial u}{\partial r}+\nu \frac u r-\beta (1+\nu)\right)\nonumber,\\
  \sigma_{\theta\theta}(r)&=&\frac E {1-\nu^2}\left(\frac u r+\nu \frac{\partial u}{\partial r}-\beta (1+\nu)\right) \ .
\end{eqnarray}
Note that by the symmetry of the problem the shear stress vanishes. The breaking of this symmetry will be important for the formation of the spiral crack patterns presented below.
The stress components have their maximum (absolute value) at the middle of the circular domain and are given by
\begin{equation}
  \label{scaling}
\sigma_{rr}(0)=\sigma_{\theta\theta}(0)=\frac {\beta E}{1-\nu}\left(\frac{1+\nu}{2I_0(\sqrt a R)}-1\right).
\end{equation}
The magnitude of the stress components monotonically increases with $R$ and in the limit $R\rightarrow \infty$ the stress components achieve the value $-E\beta/(1-\nu)$. In the limit $R=0$ the stress becomes $-E\beta/2$. From Eq. (\ref{scaling}) we can now provide an estimate of a critical domain size that will fracture under a predefined yield stress. As long as this yield stress is lower than the material stress fracture will form and grow. Depending on the material contraction the fractures may develop into spiral shaped patterns or hierarchical networks. In both cases the maximum stress is reduced by the propagating crack and only when the stress drops below the yield stress the fracturing stops. In Fig. \ref{distarea} we show a fracture network resulting from numerical solutions as explained below, formed from an initial contraction of the substrate and a predefined yield stress level. According to Eq. (\ref{scaling}) each domain division reduces the stress and an average linear size $\langle R \rangle$ of the domains can be found for a given yield stress by inverting Eq.(\ref{scaling}).

The non-uniformity of the elastic layer and the complex boundary conditions make it hard and often impossible to find an analytical solution to the displacement equation. Therefore we have implemented a numerical method based on the Galerkin finite element discretization using an adaptive triangular meshing. In the vicinity of a propagating crack tip we highly increase the resolution by decreasing locally the area of the triangular elements and thereby allow for an accurate computation of the stress intensity factors of the propagating crack. The drying process is simulated by applying a body force to the elements, i.e. we shift the equilibrium position by adding an extra force term on the right hand side of Eq. (\ref{forcebalance}). In that way we can readily add disorder into the system by selecting the magnitude of the local body force from a random distribution. In the simulations on hierarchical fracture networks presented below, we use a uniform distribution with unit mean.
\begin{figure}
\epsfig{width=.40\textwidth,angle=0,file=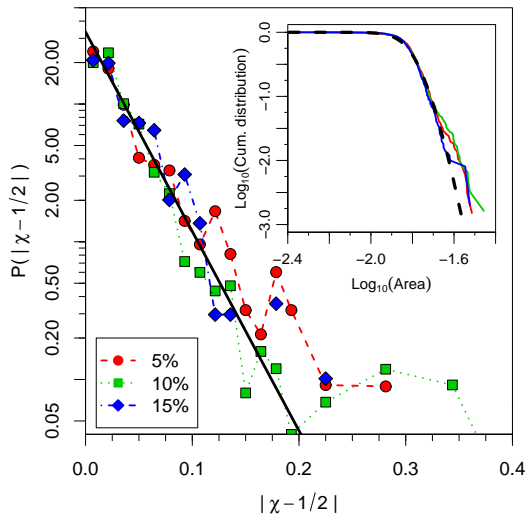}
\caption{Color online. The distribution density of $|\chi-1/2|$ for simulations with $5\%,10\%$ and $15\%$ disorder on the substrate restraining. The distributions are averaged over domains formed at the $6$th generation of cracks. No significant variation is seen at these fairly low levels of disorder. The black line on top represents a best fit with an exponential distribution $\exp(-|\chi -1/2|/\alpha)$ where $\alpha=0.03$. In the inset we show for the same data the cumulative distributions of the domain areas. The dashed line on top is an estimate of the distributions considering the individual domain divisions to be uncorrelated (see text).}\label{chidist}
\end{figure}

{\bf Crack Initiation and propagation}.  Here we present in detail how we nucleate cracks and model their evolution. First we find the points which have the highest stress and exceed the critical value. The stress is determined along the principal axes of the stress matrix  $\sigma_{ij}$, i.e. the principal stress. Whenever the yield stress is exceeded, we nucleate at the point of yielding a small semi elliptical void with an eccentricity of $0.998$. The major axis of the ellipse is aligned in the direction of the maximum principal stress. We allow the crack to evolve according to the Griffith criterion and the principle of local symmetry, i.e. the crack will grow in a direction such as to annul the local shear component at the crack tip. At each step of propagation we compute the stress in every element near the crack tip and find the \emph{stress intensity factors} from a best fit to the equations \cite{99B}.
\begin{eqnarray}
\sigma_{\theta \theta}&=&\frac{K_I}{\sqrt{2\pi r}}\cos^3\frac{\theta}{2} - 3\frac{K_{II}}{\sqrt{2\pi r}}\sin\frac{\theta}{2}\cos^2\frac{\theta}{2}\ , \\
\label{stt}
\sigma_{r \theta}&=&\frac{K_I}{\sqrt{2\pi r}}\sin\frac{\theta}{2}\cos^2\frac{\theta}{2} + \frac{K_{II}}{\sqrt{2\pi r}}\cos\frac{\theta}{2}(1-3\sin^2\frac{\theta}{2})\nonumber \ .
\end{eqnarray}
Here $r,\theta$ are local polar coordinates with respect to the crack tip with $\theta$ measured from the line following the direction of the crack. $\sigma_{\theta\theta}$ and $\sigma_{r\theta}$ are the circumferential tensile stress and the shear stress, respectively.  $K_I$ and $K_{II}$ are the unknown stress intensity factors for mode I and II, respectively.
The principle of local symmetry is satisfied if the crack grows in a direction given by an angle $\alpha$ where $K_{II}\rightarrow0$. Suppose that the crack forms an infinitesimal kink at an angle $\alpha$ from the old direction of the crack, we can define the local mode I and mode II stress intensity factors,
\begin{eqnarray}
K_I(\alpha)&=&\lim_{r\rightarrow0}\sigma_{\theta\theta}\sqrt{2\pi r}\\\nonumber
&=&K_I\cos^3\frac{\theta}{2} - 3K_{II}\sin\frac{\theta}{2}\cos^2\frac{\theta}{2}\ . \\\label{KII}
K_{II}(\alpha)&=&\lim_{r\rightarrow0}\sigma_{r\theta}\sqrt{2\pi r}\\
&=&K_I\sin\frac{\theta}{2}\cos^2\frac{\theta}{2} + K_{II}\cos\frac{\theta}{2}(1-3\sin^2\frac{\theta}{2})\ . \nonumber
\end{eqnarray}
\begin{figure}
\epsfig{width=.42\textwidth,file=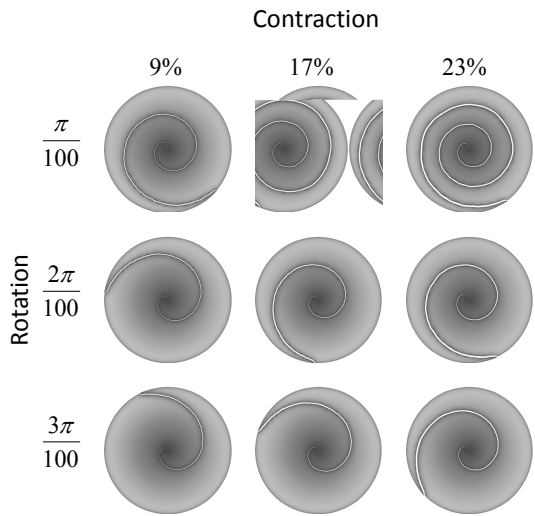}
\caption{Simulation of spiral cracks for various values of the material contraction $\beta$ and the rotation angles. In all the panels, the crack was initiated at the
center and was allowed to propagate until it reached the outer boundary. Note that the smaller the shear stress is the more pronounced is the spiralling.}
\label{spiralfig}
\end{figure}
Whether the crack propagates or not is dictated by the Griffith criterion, i.e.
the energy balance of the energy release rate into the crack tip region  must balance the dissipation involved in the crack propagation.
For a kinked crack, the energy release rate is,
\begin{equation}
G(\alpha)=(K_I^2(\alpha) + K_{II}^2(\alpha))/E \ .
\end{equation}
Since
\begin{equation}
\frac{dK_I}{d\alpha}\!\!=\!\!-\frac{3}{2}K_I\!\cos^2\frac{\alpha}{2}\!\sin\frac{\alpha}{2} - \frac{3}{2}K_{II}\cos\frac{\alpha}{2}(1-3\sin^2\frac{\alpha}{2})\\
\!\!=\!\!-\frac{3}{2}\!K_{II}\ ,
\end{equation}
the maximum of the strain energy release rate $dG(\alpha)/d\alpha\!\!=\!\!0$ is equivalent to $K_{II}(\alpha)\!\!=\!\!0$ or $dK_I(\alpha)/d\alpha=0$, thus the new direction of the crack $\alpha_0$ corresponds to the point where $K_I(\alpha_0)$ exhibits a maximum and $K_{II}(\alpha_0)=0$ \cite{86B}. Applying the latter to Eq. \eqref{KII} yields,
\begin{equation}
\alpha_0=2\arctan\left((K_I-\sqrt{K_I^2+8K_{II}^2})/4K_{II}\right) \ .
\end{equation}
We emphasize that in this model both the cracking time and the area of the fragmented elements depend only on material contraction and initial disorder. In a natural system this may not always be the case since material properties and disorder can evolve in time. Uniform contraction of the elastic layer produces homogenous hierarchical crack patterns. One can in this case argue that the effect of the crack is to partition the mother area $A$ (of generic shape) into two areas $A_1 = \chi A$ and $A_2 = (1 - \chi )A$, where $0 < \chi < 1$ is a random variable whose distribution (that must be symmetric under the transformation $\chi \mapsto 1 - \chi)$ is unknown. In Fig. \ref{chidist} is the distribution of $|\chi-1/2|=|A_1-A_2|/A$ shown together with a best fit to an exponential distribution. Although the domain areas are correlated to their mother domains, the exponential distribution of $\chi$ allows for a simple estimate of the area distribution by neglecting the correlation. That is the areas at the n'th generation level are can be determined by a product of $n$ random numbers drawn from the exponential distribution, i.e. $A_i^{(n)}=A_0\prod _{j}^n\chi_{ij}$. In Fig. \ref{chidist}, we show in the inset a distribution for domain areas at the 6th generation together with the distribution of the product of random numbers. After a few number of generations this distribution will approach a log-normal distribution. The deviation from the exponential distribution for larger values of $|\chi-1/2|$ will for an increasing number of fracture generations lead to a less good fit using the exponential distribution as an approximation.

\begin{figure}
\epsfig{file=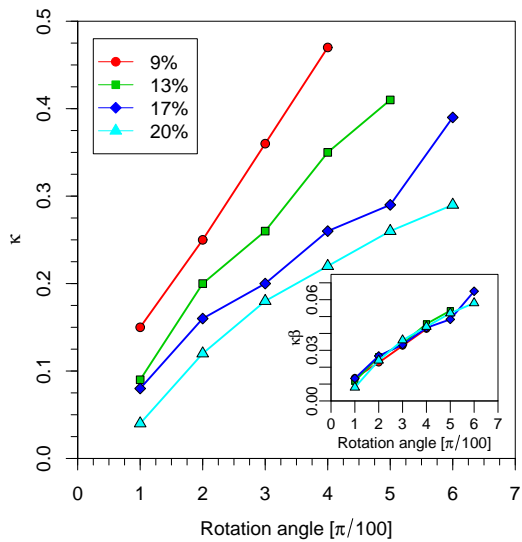,width=.41\textwidth}
\caption{Color online. The figure shows the relation between $\kappa$ and the rotation angle for four different values of the material contraction $\beta$. $\kappa$ is computed as the exponent of a best fit to a logarithmic spiral,  $r(\phi)=r_0 e^{\kappa \phi}$. The rotation angle is the prefactor $\theta_0$ used in the expression for the relative rotation between the substrate and the thin film. The inset shows a data collapse of $\kappa \beta$ for the same curves and is in agreement with the simple scaling form $\kappa \sim \theta_0/\beta $. }\label{logspiral}
\end{figure}

{\bf Spirals}. We now investigate what happens when the contraction is non-uniform and has smooth gradients. Experiments on thin films attached to a substrate by a spin-coating technique reveal a broad range of crack patterns \cite{03S} ranging from networks of cracks to single cracks spiraling outwards from their site of nucleation.  The nucleation is usually taking place at localized sites with high stress typically generated from small defects or inclusions in the material. If the material contraction is uniform in the neighborhood around the crack, the crack would propagate straight towards the material boundary where it would curve to meet the boundary at a right angle. However if the material contraction increases smoothly away from the site of nucleation, straight cracks would be unstable and would start to curve.  In that way circular shaped cracks can be formed. During the preparation of thin film coatings (such as spin-coating) it is not uncommon to have a minor residual shear stress. The shear stress breaks the symmetry of the system and the circular crack may then turn into a spiral.
If we alter the contraction such that it increases linearly away from a given site of crack nucleation, (e.g. follows a simple linear form $\beta(r)=\beta r$, and add a small shear stress by rotating the elastic layer relative to the underlying substrate with an angle $\theta_{eq}(r)=\theta_0  r^\gamma$), the crack would propagate along a spiral trajectory. Different powers of $\gamma$ result in the formation of different spirals. For the simulation of the crack propagation we use an initially circular symmetric system. A crack is then initiated at the center of the circle and is allowed to propagate according to the Griffith criterion until it reaches the boundary. The results using a rotation of the substrate by a power $\gamma=1/2$ are shown in Fig. \ref{spiralfig} using various values for the prefactors $\theta_0$ and $\beta$, respectively.
The cracks have a shape that fit well a logarithmic spiral, i.e. they have a form $r(\theta) = A\exp (\kappa \theta)$ where $\kappa$ depends on the material contraction $\beta$ and the rotation $\theta_0$. In Fig. \ref{logspiral} we show best fits of $\kappa$ as function of $\theta_0$ and for four values of $\beta$.

In summary, we pointed out the role of residual stresses in determining the crack patterns in drying thin substrates. For spiral patterns we
related the properties of the spiral to the degree of residual shear stress left in the layer. For hierarchical patterns we determined the position
where new cracks initiate as a function of the mother cell, and offered a relation of final
mean size of cells to the critical value of the yield stress.

\acknowledgments
We thank M. Adda-Bedia for proposing the study of spiral cracks. This work has been supported by the \textsl{PGP}, a Center of Excellence at the University of Oslo, the German Israeli Foundation and the Minerva Foundation, Munich, Germany.


\begin{thebibliography}{99}
\bibitem{88SM} A. T. Skjeltorp and P. Meakin, Nature {\bf 335}, 424 (1988).

\bibitem{05BPAAC} S.~Bohn, J.~Platkiewicz, B.~Andreotti, M.~Adda-Bedia, and Y.~Couder, Phys.Rev.R {\bf 71}, 046215 (2005)

\bibitem{05B2}
S. Bohn, S. Douady and Y. Couder, {Phys. Rev. Lett.}, \textbf{94}, 054503
(2005).

\bibitem{94G}
A. Groisman and E. Kaplan,  {Europhys. Lett.}, \textbf{ 25}, 415
(1994).

\bibitem{08IJMMF} K. Iyer, B. Jamtveit, J. Mathiesen, A. Malthe-Sørenssen and J. Feder, EPSL {\bf 267}, 503 (2008).

\bibitem{03S}
M. Sendova and K. Willis, {Appl. phys. A}, \textbf{76}, 957 (2003).


\bibitem{00M}
J. Malzbender and G. de With, {Thin Solid Films} \textbf{359}, 210 (2000).

\bibitem{07K}
E. Katzav, M. Adda-Bedia and B. Derrida, {Europhys. Lett.},
\textbf{ 78}, 46006 (2007).

\bibitem{07S}
S. Sadhukhan, S. R. Majumder, D. Mal, T. Dutta and S. Tarafdar,
{J. Phys. Cond. Matt.}, \textbf{19}, 356206 (2007).

\bibitem{98A}
J. V. Andersen and L. J. Lewis, {Phys. Rev. E} \textbf{57}, R1211 (1998)

\bibitem{97B}
A. Buchel and J. P. Sethna, {Phys. Rev. E} \textbf{55}, R7669 (1997)

\bibitem{03L}
J. Liang, R. Huang, J. H. Prevost and Z. Suo, {Int. J. Solids Struct.}
\textbf{40}, 2343 (2003).

\bibitem{86B}
D. Broek, \textit{Elementary engineering fracture mechanics}. Kluwer Academic Publishers, Dordrecht, 1986.

\bibitem{99B}
K. B. Broberg, \textit{Crack and Fracture}, (Academic Press, London) 1999.


\end{thebibliography}
\end{document}